\documentclass[twocolumn,times]{aastex63}

\usepackage{newtxtext,newtxmath}
\usepackage{graphicx}	
\usepackage{amsmath}	
\usepackage{amssymb}	
\usepackage{mathrsfs}

\received{}
\revised{}
\accepted{}
\submitjournal{ApJL}

\shorttitle{FRB Breakouts from Magnetar Burst Fireballs}
\shortauthors{Ioka}

\begin{document}

\title{Fast Radio Burst Breakouts from Magnetar Burst Fireballs}

\correspondingauthor{Kunihito Ioka}
\email{kunihito.ioka@yukawa.kyoto-u.ac.jp}

\author[0000-0002-3517-1956]{Kunihito Ioka}
\affiliation{Center for Gravitational Physics, Yukawa Institute for Theoretical Physics, Kyoto University, Kyoto 606-8502, Japan}




\begin{abstract}
  The recent discovery of a Mega-Jansky radio burst occurring simultaneously with short X-ray bursts from the Galactic magnetar (strongly magnetized neutron star (NS)) SGR 1935+2154 is a smoking gun for the hypothesis that some cosmological fast radio bursts (FRBs) arise from magnetar bursts. We argue that the X-ray bursts with high temperature $T \gtrsim 30$ keV entail an electron--positron ($e^{\pm}$) outflow from a trapped--expanding fireball, polluting the NS magnetosphere before the FRB emission. The $e^{\pm}$ outflow is opaque to FRB photons, and is strongly Compton-dragged by the X-ray bursts. Nevertheless, the FRB photons can break out of the $e^{\pm}$ outflow with radiation forces if the FRB emission radius is larger than a few tens of NS radii. A FRB is choked if the FRB is weaker or the X-ray bursts are stronger, possibly explaining why there are no FRBs with giant flares and no detectable X-ray bursts with weak FRBs. We also speculate that the $e^{\pm}$ outflow may be inevitable for FRBs, solving the problem of why the FRBs occur only with high-$T$ X-ray bursts. The breakout physics is important for constraining the emission mechanism and electromagnetic counterparts to future FRBs.
\end{abstract}



\section{Introduction}

Fast radio bursts (FRBs) are enigmatic radio transients
with extremely high brightness temperature $T_b \sim 10^{35}$ K
\citep{Lorimer+07,Thornton+13,Katz18,Cordes19,Petroff19}.
New clues are being found such as
repeating FRBs \citep{Spitler+16},
periodic FRBs \citep{FRB180916,Rajwade+20,Ioka+20}
\citep[see also][]{Grossan20},
and so on.
Regardless of their origin,
they are also unique probes for cosmology \citep{Ioka03,Inoue04},
with actual observations being analyzed \citep{Macquart+20}.

Recently, a smoking gun has been discovered with
the detection of Mega-Jansky FRB 200428 \citep{CHIME/FRB+200428,Bochenek+20}:
two radio pulses temporally coincide with short X-ray bursts from
the magnetar SGR 1935+2154 in our Galaxy
\citep{Li+20,Mereghetti+20,Ridnaia+20,Tavani+20}.
The energy is $\sim 40$ times smaller than the faintest extragalactic FRBs,
but three orders of magnitude larger than the brightest giant radio pulses
from Galactic neutron stars (NSs).
Therefore, it is fair to say that magnetar bursts can produce FRBs
\citep[as widely suspected; see, e.g.,][]{Popov+10,Kulkarni+14,Lyubarsky14,Pen+15,Cordes+16,Katz16,Murase+16,Kashiyama+17,Beloborodov17,Kumar+17,Metzger+17,Wadiasingh+19,Ioka+20}.

At the same time, however, new puzzles also arise.
No FRB is associated with other X-ray bursts
down to eight orders of magnitude fainter than FRB 200428
\citep{Lin+20}.
An apparent difference of FRB 200428
is the cutoff energy of the spectrum ($T_{\rm cut} \sim 80$ keV),
which is higher than that of other X-ray bursts from SGR 1935+2154
\citep[$T_{\rm cut} \sim 10$ keV;][]{Li+20,Lin+20b,Lin+20c,Ridnaia+20,Younes+20}.
Weaker radio bursts without X-ray bursts are also detected
with $112\pm 22$ Jy ms and $24\pm 5$ Jy ms separated by 1.4 s
\citep{Kirsten+20},
and with $60$ mJy ms \citep{Zhang+20}
like previously known radio pulses from magnetars
\citep[][]{Camilo+06,Levin+10,Shannon+13,Eatough+13,Esposito+20}.
The emission region remains controversial
\citep{Lu+20,Lyutikov+200428,Katz+20,Margalit+20,Yu+20,Yuan+20},
whether it is in the magnetosphere of the NS
\citep{Kashiyama+13,Cordes+16,Lyutikov+16,Kumar+17,Zhang17,Yang+18,Lyubarsky20,Kumar+20,Ioka+20}
or far away at the circumstellar matter
interacting with relativistic ejecta from the NS
\citep{Lyubarsky14,Murase+16,Waxman+17,Beloborodov17,Metzger+17}
\citep[see also][]{Melrose+06,Lu_Kumar18,Wadiasingh+19,Lyutikov20}.

\begin{figure}
  \begin{center}
    \includegraphics[width=\linewidth]{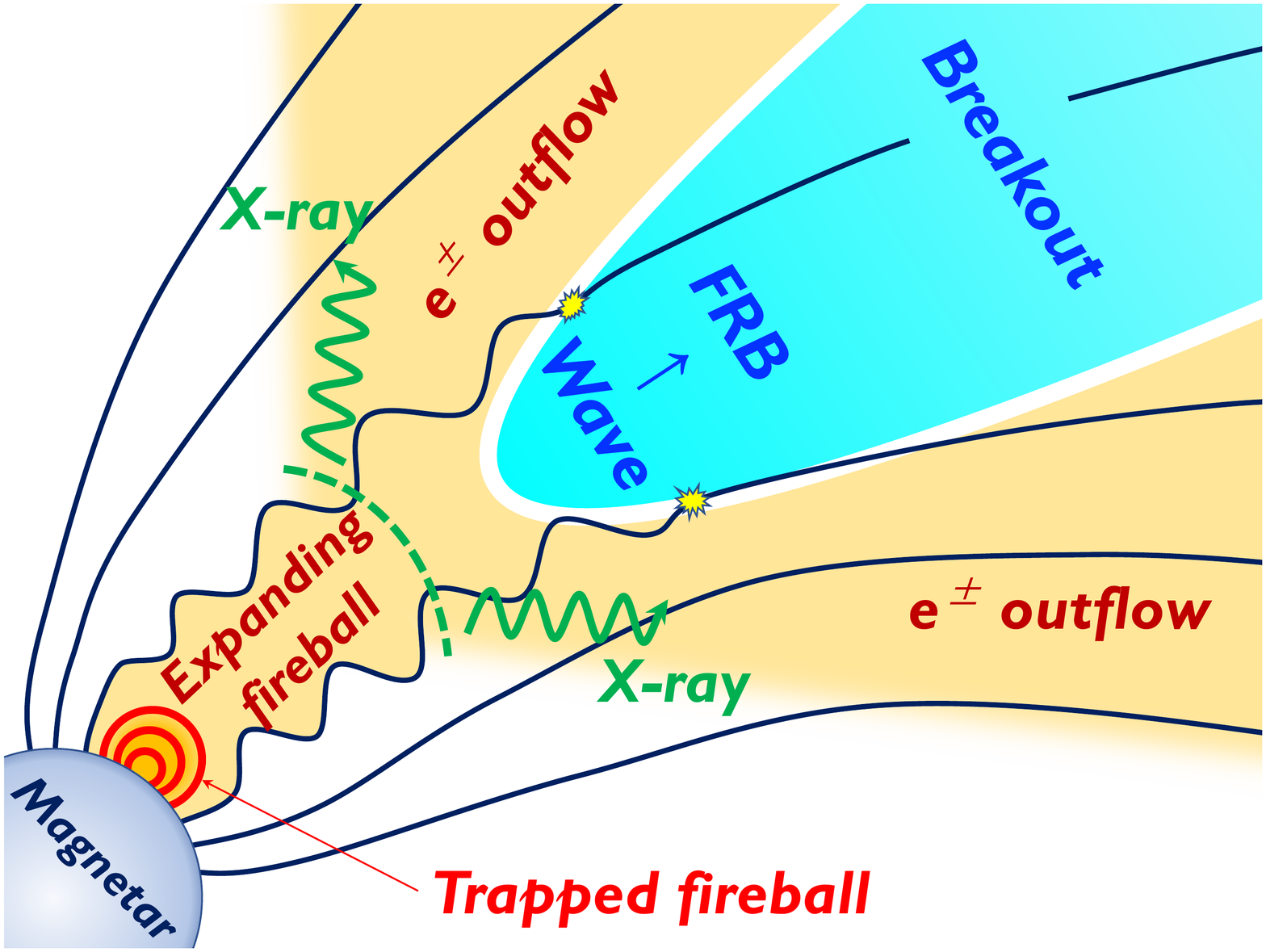}
  \end{center}
  \caption{
    Schematic configuration.
    Energy is released near the NS surface,
    leading to a trapped fireball of $e^{\pm}$ and X-rays
    in the closed magnetic field line,
    and to magnetohydrodynamic (MHD) waves along the large-scale field line,
    which dissipate into FRB photons
    at a distance more than a few tens of NS radii.
    X-rays from the trapped fireball
    create an expanding fireball,
    which first propagates along the large-scale magnetic tube
    and then diffuses across the field line.
    Accordingly, the expanding fireball releases X-rays and $e^{\pm}$ outflow.
    The $e^{\pm}$ outflow is thick to induced Compton scatterings of FRB photons.
    The FRB photons break out of the precursory $e^{\pm}$ outflow
    with radiation forces.
  }
  \label{fig:model}
\end{figure}

In this Letter, we suggest that
the X-ray bursts with $T_{\rm cut} \sim 80$ keV
entail electron--positron ($e^{\pm}$) outflows,
and FRB photons, if emitted in the magnetosphere,
penetrate and break it out with radiation forces that can be observed as FRBs,
as in Fig.~\ref{fig:model}.
In Sec.~\ref{sec:TEFB},
we examine a trapped fireball for the X-ray bursts
and show that it is connected to an expanding fireball,
leading to an $e^{\pm}$ outflow,
because $T_{\rm cut} \sim 80$ keV
is high enough to create abundant $e^{\pm}$
outside the trapped fireball.
In Sec.~\ref{sec:prop&breakout},
we discuss that the $e^{\pm}$ outflow is
optically thick to induced Compton scatterings
of FRB photons unless the photons are extremely beamed,
and obtain the breakout condition,
taking the Compton drag on the $e^{\pm}$ outflow
by the X-ray bursts into account.
This limits the emission radius larger than
a few tens of NS radii.
In Sec.~\ref{sec:summary},
we discuss implications for the above puzzles.
We use $Q_{,x}\equiv Q/10^x$ in cgs units
with the Boltzmann constant $k_B=1$.


\section{Trapped--Expanding Fireball}\label{sec:TEFB}

The magnetar SGR 1935+2154 has a period $P = 3.24$ s and
a period derivative ${\dot P} = 1.43 \times 10^{-11}$ s s$^{-1}$.
We estimate the magnetic field at the pole
\begin{eqnarray}
  B_p \sim 2 \times 10^{14}\,{\rm G}\, B_{p,14.3},
\end{eqnarray}
the light cylinder radius $r_{L}={cP}/{2\pi} \sim 2 \times 10^{10}$ cm,
and the spin-down luminosity $L_{\rm sd} \sim 2 \times 10^{34}$ erg s$^{-1}$,
where $R=10^{6}$ cm is the NS radius.

We consider a sudden, localized energy release near the NS surface
via crust cracking or magnetic reconnection (see Fig.~\ref{fig:model}).
The energy dissipated in the closed field line
forms a trapped fireball of $e^{\pm}$ and X-rays,
powering the X-ray bursts.
The energy also propagates along a large-scale field line
as magnetohydrodynamic (MHD) waves
and dissipates into coherent radio waves as FRBs
far away from the NS surface
\citep{Lyubarsky20,Kumar+20}.
Note that the trapped fireball is the standard model for
soft gamma repeater (SGR) bursts
\citep{Thompson+95,Thompson+01,Yang+15},
which naturally explains the longer timescale
than the crossing time $\ell_X/c \sim 3\times 10^{-7}$ s,
such as
the delay time of the X-ray peak from the FRB pulse
\citep[$\sim 6.5 \pm 1.0$ msec;][]{Mereghetti+20}
and the X-ray peak widths \citep[$\sim 3$ msec;][see Sec.~\ref{sec:summary}]{Li+20}.
The observed unusual spectrum
\citep{Younes+20,Ridnaia+20,Li+20}
may be explained by
a different configuration of magnetic fields (see Sec.~\ref{sec:TFB}).
A similar setup of an active region connected to
  high quasi-polar altitudes is also considered by \citet{Younes+20}.

The onset of the X-ray bursts
starts $\sim 30$ msec before the FRBs,
and the hardness ratio also rises with the flux
\citep{Li+20,Mereghetti+20}.
This is followed by the temporally correlated FRBs and X-ray peaks,
suggesting that
the energy is generated at the same place.

As shown below,
an expanding fireball of $e^{\pm}$ and X-rays is also
launched from the trapped fireball
because of the high cutoff energy $T_{\rm cut}\sim 80$ keV.
The high-energy tail of the X-rays
exceeds the pair threshold and creates abundant $e^{\pm}$ pairs
outside the trapped fireball, which are highly opaque.
The X-rays should be carried with the $e^{\pm}$
along the large-scale field,
and released at a large distance for the X-ray bursts.
Because the X-ray onset begins before the FRBs,
the precursory $e^{\pm}$ outflow is widely distributed
along the magnetic field line,
and the FRB emission is likely affected by the $e^{\pm}$ outflow
(see Sec.~\ref{sec:prop&breakout}).\footnote{
  This is not the case if the energy is transferred through the NS crust
  and released far away from the trapped fireball
  \citep{Lu+20}.
  In this case, the more spread out, the less energy there is.
}

In this section, we model the trapped--expanding fireball
associated with the X-ray bursts.
%
X-rays and $e^{\pm}$ are released after several steps.
(i) X-rays are emitted from the trapped fireball.
(ii) $e^{\pm}$ are created outside the trapped fireball and
the fireball flows along the large-scale magnetic field.
(iii) X-rays diffuse out transversely from the $e^{\pm}$ and associated magnetic field line,
creating $e^{\pm}$ in a wide range of the surrounding magnetic field lines.
(iv) X-rays are released, and pair annihilation is frozen.
We obtain the resulting density and Lorentz factor of the $e^{\pm}$ outflow,
taking the Compton drag by X-rays into account.

\subsection{Trapped Fireball}
\label{sec:TFB}

The size of the trapped fireball is estimated from
the X-ray luminosity $L_X \sim 10^{41}$ erg s$^{-1}\, L_{X,41}$
and cutoff energy,
which is identified\footnote{
  Non-thermalization should happen later,
  at least softening the low-energy spectral index $\alpha$ as observed.
  The cutoff energy may be also shifted,
  by photon splitting, resonant scattering, and so on,
  but we do not discuss this here.
  Note that the spectral peak energy is about
    $(\alpha+2) T_{\rm cut} \sim 37$ keV for $\alpha\sim -1.5$
    \citep{Lin+20c,Younes+20}.
}
with the effective temperature of the trapped fireball
$T = T_{\rm cut} \sim 80$ keV $T_{1.9}$, as
\begin{eqnarray}
  \ell_X \sim \left(\frac{L_X}{2 \pi c a T^4}\right)^{1/2}
  \sim 1\times 10^{4}\,{\rm cm}\, L_{X,41}^{1/2} T_{1.9}^{-2},
  \label{eq:lx}
\end{eqnarray}
where $a$ is the radiation constant.
This is much smaller than the NS radius,
implying non-dipole magnetic structure.
The magnetic energy in the trapped fireball is
$(2\pi/3) \ell_X^3 (B^2/8\pi) \sim 10^{40}$ erg $B_{14.3}^2 \ell_{X,4}^3$,
which can confine the burst energy
for the observed duration $\sim 0.1$ s.\footnote{
  The energy injection into the trapped fireball may not be a one-shot,
  and/or several trapped fireballs may be created,
  as suggested by the multiple X-ray peaks.
  However, stationarity is not a bad approximation because
  the luminosity is constant within a factor of a few.
  Note that the NS rotates by $\sim 2\pi/100$ radian
  during $\sim 30$ ms between the peaks,
  which is negligible for the nearly isotropic X-ray emission.
}
There is a temperature gradient inside the trapped fireball
that realizes the energy transfer consistent with the X-ray luminosity
\citep{Lyubarsky02}.

\begin{figure}
  \begin{center}
    \includegraphics[width=\linewidth]{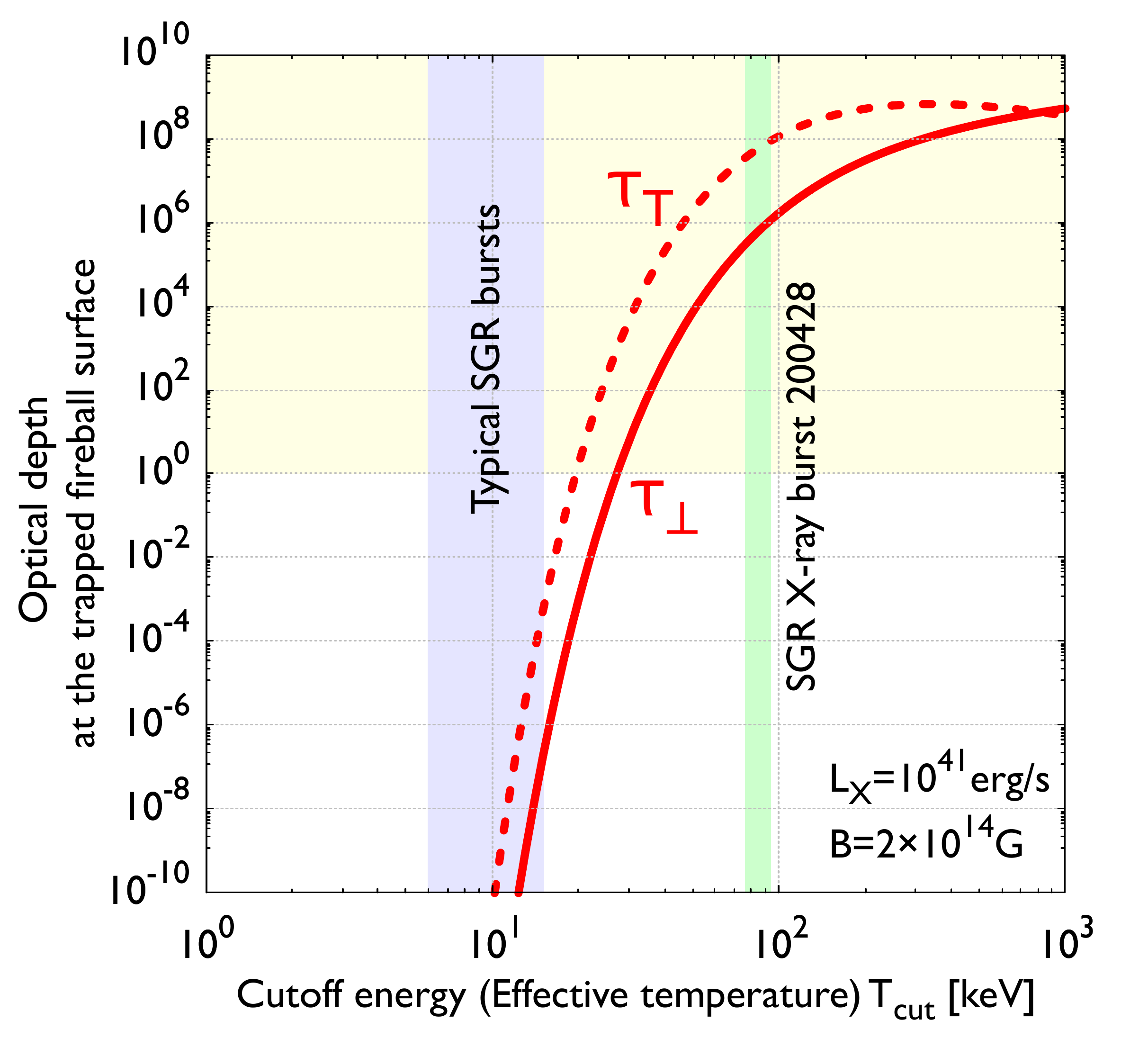}
  \end{center}
  \caption{
    Optical depth at the surface of the trapped fireball,
    $\tau_{\perp}$ (for E-mode) and $\tau_T$ (for O-mode),
    as a function of the cutoff energy $T_{\rm cut}$
    for $L_{X}=10^{41}$ erg s$^{-1}$ and $B=2\times 10^{14}$ G.
    It is optically thick even outside the trapped fireball
    for the SGR X-ray bursts associated with FRB 200428
    because $T_{\rm cut}$ is higher than that of typical bursts
    and the high-energy tail of the X-rays
    above the pair threshold creates
    abundant $e^{\pm}$.
    In order for the X-rays to be observed,
    the trapped fireball should launch an expanding fireball.
  }
  \label{fig:tau}
\end{figure}

In this event, the cutoff energy $T_{\rm cut}$ is much higher than typical
\citep{Li+20,Ridnaia+20,Younes+20}.
Even the outside of the trapped fireball is found to be optically thick
(inside a photosphere).
The equilibrium number density of $e^{\pm}$ produced by
the high-energy tail of X-rays from the trapped fireball is
\begin{eqnarray}
  n_{\pm}=\frac{e B m_e}{(2\pi^3)^{1/2}\hbar^2}
  \left(\frac{T}{m_e c^2}\right)^{1/2}
  \exp\left(-\frac{m_e c^2}{T}\right),
  \label{eq:npm}
\end{eqnarray}
where the effective temperature $T=T_{\rm cut}$ is less than
the excitation energy of the first Landau level for electrons
$h \nu_B=(m_e^2 c^4+2 \hbar c eB)^{1/2}-m_e c^2$
\citep{Thompson+95}.
The Rosseland mean optical depth of a photon with electric vector perpendicular to $B$
(the extraordinary mode or E-mode)
is estimated as \citep{Meszaros92,Thompson+95,Lyubarsky02}
\begin{eqnarray}
  \tau_{\perp}=\frac{4\pi^2}{5} \sigma_T
  \left(\frac{T}{m_e c^2}\frac{B_Q}{B}\right)^2 n_{\pm} \ell_X,
\end{eqnarray}
where
$B_Q={m_e^2 c^3}/{\hbar e}=4.4 \times 10^{13}\,{\rm G}$.
The orthogonal polarization state (the ordinary mode or O-mode) has
a higher optical depth
$\tau_T \sim n_{\pm} \sigma_T \ell_X$.
As shown in Fig.~\ref{fig:tau},
the outside of the trapped fireball is opaque in this event
with $T\sim 80$ keV,
while it is thin in typical bursts with $T \sim 10$ keV.
This is a critical difference from usual bursts.


Then the trapped fireball should be located at the base of
an open magnetic field.
Otherwise, if the trapped fireball is surrounded by a closed field,
the released X-rays just increase the size of the trapped fireball,
leading to a lower temperature (like typical bursts) than
that of the observation.
Therefore, for the high $T_{\rm cut}$ to be observed,
the $e^{\pm}\gamma$ plasma should expand
along the large-scale open field lines outside the trapped fireball,\footnote{
  The large-scale field is not necessarily open to infinity.
}
and finally become optically thin, keeping the observed $T_{\rm cut}$
like an expanding fireball for gamma-ray bursts
\citep{Goodman86,Paczynski86,Meszaros+00}.
In this picture, the spectral difference from typical bursts
is attributed to the magnetic field configuration (open or closed)
around the initial trapped fireball.

\subsection{Expanding Fireball along the Large-scale Magnetic Field}

The expanding fireball arising from the trapped fireball
runs along the magnetic field
because the magnetic field pressure is stronger than the fireball pressure
and the $e^{\pm}$ are frozen in the field lines.
Within a distance less than the NS radius $r<R$,
the magnetic field lines do not spread that much.
The fireball moves in a tube with a nearly constant cross section,
thereby with a constant velocity (no acceleration),
constant density, and constant temperature.


At $r>R$, the magnetic field lines begin to open.
For a dipolar field, a perpendicular width expands as
\begin{eqnarray}
  \ell_{\perp}=\ell_X (r/R)^{3/2}.
\end{eqnarray}
Accordingly, the Lorentz factor and comoving temperature
of the expanding fireball evolve as \citep{Meszaros+00,Thompson+01}
\begin{eqnarray}
  \Gamma \sim (r/R)^{3/2},\quad
  T' \sim T_{\rm cut} (r/R)^{-3/2}.
  \label{eq:GammaT}
\end{eqnarray}


\subsection{$e^{\pm}\gamma$ Diffusion across the Large-scale Magnetic Field}

X-rays diffuse in the $e^{\pm}$ flow.
E-mode photons scatter less than O-mode photons.
X-rays first diffuse into the perpendicular direction
to the outflow motion, i.e., across the magnetic field.
As the temperature $T'$ drops due to expansion in Eq.~(\ref{eq:GammaT}),
the comoving $e^{\pm}$ density $n'_{\pm}$ decreases exponentially
in Eq.~(\ref{eq:npm}),
and eventually the diffusion time of the E-mode photons
becomes less than the dynamical time
\begin{eqnarray}
  t'_{\rm diff} \equiv \frac{\ell_{\perp}}{c} \tau_{\perp} < \frac{r}{c\Gamma}
  \equiv t'_{\rm dyn},
  \label{eq:diffuse}
\end{eqnarray}
at a radius and a Lorentz factor
\begin{eqnarray}
  r=r_d \sim 1.9 R, \quad \Gamma = \Gamma_d \sim 2.6,
\end{eqnarray}
respectively.
Here we assume a dipole $B \propto r^{-3}$.
Note that the magnetic field strength $B$ and
the perpendicular width $\ell_{\perp}$ are
frame-independent as the flow motion is parallel to $B$.


The diffusing X-rays (more precisely the high-energy tail above the pair threshold) create $e^{\pm}$ pairs
outside the initial magnetic field lines.
Once the diffusion starts (i.e., $\ell_{\perp}$ expands),
the above condition in Eq.~(\ref{eq:diffuse}) is always satisfied
because the isotropic luminosity $L_{{\rm iso},X} \sim (r_d/\ell_{\perp})^2 L_X$
and the corresponding temperature $T' \sim (L_{{\rm iso},X}/2\pi r_d^2 c a \Gamma^2)^{1/4}$
decreases.
As the width expands to
$\ell_{\perp}(r_d) \sim 3.6 \times 10^4\,{\rm cm}$
($T'\sim 26$ keV) due to diffusion,
the perpendicular direction becomes optically thin to E-mode photons
$\tau_{\perp} \sim 1$.
As $\ell_{\perp}(r_d) \sim 4.8 \times 10^4$ cm
($T'\sim 22$ keV),
it also becomes thin to O-mode photons
$\tau_{T} = n'_{\pm} \sigma_T \ell_{\perp} \sim 1$.
As $\ell_{\perp}(r_d) \sim 6.0 \times 10^4$ cm
($T'\sim 20$ keV),
it also becomes thin to the radial direction
$\tau_{T} = n'_{\pm} \sigma_T r_d/\Gamma_d \sim 1$.
Then the $e^{\pm}$ creation across the magnetic field becomes ineffective.
The width of the $e^{\pm}$ outflow becomes roughly
\begin{eqnarray}
  \ell_{\perp}(r) \sim 2 \times 10^{4}\,{\rm cm}\,r_6^{3/2},
  \label{eq:lperp}
\end{eqnarray}
which is wider than the initial size $\ell_X$ in Eq.~(\ref{eq:lx}),
and it extends to $\ell_{\perp} \sim r$ at $r\sim 10^{9}$ cm.
The X-rays are released to an opening angle $\sim 1/\Gamma_d \sim 0.4$
at this stage.\footnote{
    Radiative transfer brings a factor of two anisotropy within the beaming cone
  \citep{vanPutten+16}.
}


\subsection{$e^{\pm}$ Outflow Compton-dragged by the X-Ray Bursts}

Once X-rays diffuse out to the perpendicular direction,
the equilibrium $e^{\pm}$ density drops rapidly until
the annihilation stops and their number freezes.
The relic number density is determined by the condition
that the annihilation time
$\sim 1/n'_{\pm}(r_d) \sigma(\bar \beta'_{\pm}) c \bar \beta'_{\pm}$
equals to the dynamical time $\sim r_d/c \Gamma_d$ as
\begin{eqnarray}
  n'_{\pm}(r_d) \sim \frac{\Gamma_d}{\sigma_T r_d}
  \sim \frac{\Gamma_d^{1/3}}{\sigma_T R}
  \sim 2 \times 10^{18}\,{\rm cm}^{-3}\,\Gamma_{d,0.4}^{1/3},
\end{eqnarray}
where we use Eq.~(\ref{eq:GammaT}) and 
the cross section for annihilation
$\sigma(\bar \beta'_{\pm}) \sim \sigma_T/\bar \beta'_{\pm}$
for a small thermal velocity $\bar \beta'_{\pm} \ll 1$.
Beyond the diffusion radius $r_d$,
the number density evolves as
\begin{eqnarray}
  n'_{\pm}(r) \sim \frac{\Gamma_d^{1/3}}{\sigma_T R}
  \frac{\Gamma_d}{\Gamma_{\pm}}
  \left(\frac{r}{r_d}\right)^{-3}
  \sim 3 \times 10^{16}\,{\rm cm}^{-3}\,\Gamma_{d,0.4}^{10/3} \Gamma_{\pm}^{-1} r_7^{-3},
  \label{eq:npm(r)}
\end{eqnarray}
where $\Gamma_{\pm}$ is the Lorentz factor of the $e^{\pm}$ outflow,
because the number is conserved and
the perpendicular width of the outflow follows Eq.~(\ref{eq:lperp}).
This is
$\sim 10^{7}$ times
larger than the Goldreich--Julian density.

The released X-rays make cyclotron resonant scatterings
\citep{Canuto+71,Thompson+02}
at a radius around
\begin{eqnarray}
  r_{\rm res} \sim R \left(\frac{e B_p}{2\pi m_e c \nu}\right)^{1/3}
  \sim 10^{7}\,{\rm cm}\, B_{p,14}^{1/3} \nu_{\rm keV}^{-1/3},
\end{eqnarray}
although the Thomson optical depth is below unity
\begin{eqnarray}
\tau_T \sim n'_{\pm} \sigma_T {r}/{\Gamma_{\pm}} \sim 
0.2\, \Gamma_{d,0.4}^{10/3} \Gamma_{\pm}^{-2} r_7^{-2}.
\label{eq:Thomson}
\end{eqnarray}
The X-ray field is basically isotropized within this radius.
An X-ray pulse is also delayed and broaden by the crossing time
$\sim 2 r_{\rm res}/c \sim 1$ ms.
The observed delay
\citep[$\sim 6.5\pm 1.0$ ms;][]{Mereghetti+20}
and width \citep[$\sim 3$ ms;][]{Li+20}
of the X-ray bursts
are larger than this timescale,
implying the trapping to the fireball.


The Lorentz factor $\Gamma_{\pm}$ or velocity $c\beta_{\pm}$
of the $e^{\pm}$ outflow
is basically determined by the Compton drag due to X-rays.
Given the X-ray energy density $u'_{X}=L_{X}/4\pi r^2 c \Gamma_{\pm}^2$,
the Compton drag time $t'_{\rm dr}=m_e c^2/c\sigma_T u'_{X}$
is less than the dynamical time $t'_{\rm dyn}=r/c\Gamma_{\pm}$ if 
\begin{eqnarray}
  \Gamma_{\pm} < \left(\frac{L_{X} \sigma_T}{4\pi m_e c^3 r}\right)^{1/3}
  \sim 30\,L_{X,41}^{1/3} r_{7}^{-1/3}.
  \label{eq:drag}
\end{eqnarray}
Thus, in the magnetosphere
the Compton drag is basically very strong
due to the strong X-ray emission
\citep{Yamasaki+20}.
The velocity of the $e^{\pm}$ outflow is forced to be
\begin{eqnarray}
  \beta_{\pm}=\cos \theta_{kB},
  \label{eq:beta}
\end{eqnarray}
when the photons stream at an angle $\theta_{kB}$
with respect to $B$ \citep{Thompson+02,Beloborodov13,Yamasaki+20}.
Within $r<r_{\rm res}$,
the X-ray field is nearly isotropic and hence
$\Gamma_{\pm} \sim 1$.
At $r\gg r_{\rm res}$, X-rays travel radially,
and $\tan \theta_{kB} = (1/2) \tan \theta$
because a dipole field line satisfies
$\sin^2\theta/r ={\rm const.}$,
where $\theta$ is a polar angle.
Therefore, the $e^{\pm}$ outflow is mildly relativistic
except for the polar region.
Note that $\Gamma_{\pm} \sim \theta_{kB}^{-1} \sim 2/\theta$
for $\theta_{kB} \ll 1$.
Note also that the above is the most simplistic argument
  and do not account for strong angle dependence of
  resonant scattering or its kinematics.

In the polar region $\theta \ll 1$,
the acceleration of the $e^{\pm}$ outflow is limited by
$\Gamma=r/r_{\rm res}$ like an expanding fireball
because this is the frame in which the X-ray field is isotropic
\citep[e.g.,][]{Meszaros+00}.
Then the Compton drag is effective ($t'_{\rm dr}<t'_{\rm dyn}$) up to
\begin{eqnarray}
  \Gamma_*=\left(\frac{L_{X} \sigma_T}{4\pi m_e c^3 r_{\rm res}}\right)^{1/4}
  \sim 10\, L_{X,41}^{1/4} r_{{\rm res},7}^{-1/4}.
  \label{eq:G*}
\end{eqnarray}

Given the density in Eq.~(\ref{eq:npm(r)}) and velocity in Eq.~(\ref{eq:beta}),
the isotropic kinetic luminosity of the $e^{\pm}$ outflow is obtained as
\begin{eqnarray}
  L_{\pm} &\sim& 4\pi r^2 n'_{\pm}(r) m_e c^3 \beta_{\pm}^3 \Gamma_{\pm}^2
  \sim \frac{4\pi R m_e c^3}{\sigma_T} \beta_{\pm}^3 \Gamma_{\pm} \Gamma_d^{10/3}
  \left(\frac{r}{R}\right)^{-1}
  \nonumber\\
  &\sim& 1 \times 10^{36}\,{\rm erg}\,{\rm s}^{-1}\,
  \beta_{\pm}^3 \Gamma_{\pm} \Gamma_{d,0.4}^{10/3} r_7^{-1},
  \label{eq:Lpm}
\end{eqnarray}
which is much weaker than the X-ray ($\sim 10^{41}$ erg s$^{-1}$) and
FRB ($\sim 10^{38}$ erg s$^{-1}$).
Along the open field line, the kinetic luminosity may be
comparable to the spin-down luminosity at the light cylinder.

\section{Propagation and Breakout of FRB}\label{sec:prop&breakout}

The $e^{\pm}$ outflow from the trapped--expanding fireball
is an obstacle for FRB photons to propagate in the magnetosphere.
In Sec.~\ref{sec:prop}, we show that it is generally optically thick to
induced Compton scatterings of FRB photons
\citep{Wilson+78,Thompson+94,Lyubarsky08}
because the brightness temperature of the FRB is extremely high
($T_{b} \sim 10^{33}$ K)
and the scattering cross section is enhanced by
the occupation number of photon quantum states
$T_{b}/h\nu \sim 10^{34}\, T_{b,33} \nu_9^{-1}$.
Therefore, the FRB photons should break out of the $e^{\pm}$ outflow
in order to be observed.
In Sec.~\ref{sec:breakout}, we obtain the breakout condition,
where the Compton drag on the $e^{\pm}$ outflow by the X-rays
is essential.
Radiation forces of FRB photons are also considered by
\citet{Kumar_Lu20},
particularly for restricting the far-away FRB models.

In this Letter, we do not discuss the generation of coherent radio photons.
We assume that the FRB photons are generated,
and solely discuss whether the photons can propagate and break out of
the $e^{\pm}$ outflow associated with the X-ray bursts
\citep[see][for other constraints]{Melrose+06}.
The physical condition of the FRB generation site
is uncertain and probably different from
that of the surrounding $e^{\pm}$ outflow
because the MHD waves with larger energy
would modify the $e^{\pm}$ outflow.

\subsection{Induced Compton scatterings by the $e^{\pm}$ outflow}
\label{sec:prop}

In the $e^{\pm}$ outflow with the number density in Eq.~(\ref{eq:npm(r)}),
the optical depth to induced Compton scatterings is very large,
\begin{eqnarray}
  \tau_{C} &\sim& \frac{3\sigma_T}{32\pi^2}
  \frac{n_{\pm}(r) L_{\rm FRB} c \Delta t_{\rm FRB}}{r^2 m_e \nu^3}
  \nonumber\\
  &\sim& 6\times 10^{21}\,
  \Gamma_{d,0.4}^{10/3} (L_{\rm FRB}\Delta t_{\rm FRB})_{35} \nu_9^{-3} r_{7}^{-5},
\end{eqnarray}
where $L_{\rm FRB} \Delta t_{\rm FRB}$ is the isotropic FRB energy,
and we assume that the outflow is non-relativistic
due to the Compton drag by the X-ray bursts in Eq.~(\ref{eq:beta}).\footnote{
  We also assume that the opening angle
  of the FRB photon beam satisfies $\theta_b > (2 c \Delta t_{\rm FRB}/r)^{1/2}$
  \citep{Lyubarsky08}.
  We also neglect the acceleration of the $e^{\pm}$ to a Lorentz factor
  comparable to the dimensionless wave strength
  \begin{eqnarray}
   a=\frac{eE_{\rm FRB}}{m_e \omega c}
   =\frac{e}{m_e \omega c}\left(\frac{2 L_{\rm FRB}}{c r^2}\right)^{1/2}
   \sim 4\times 10^{4}\, L_{{\rm FRB},38.6}^{1/2} \nu_{9}^{-1} r_{7}^{-1}.
   \nonumber
  \end{eqnarray}
}
If the $e^{\pm}$ outflow is relativistic (e.g., in the polar region),
we should make Lorentz transformations \citep[see][]{Ioka+20}.
Note that even without the $e^{\pm}$ outflow,
the system is optically thick due to the Goldreich--Julian density.
Induced Raman scatterings may be also effective.

The optical depth to the induced Compton scatterings is suppressed by
a factor $\sim \min\left[\theta_E^{-2},(\nu_B/\nu)^2\right]$
if the magnetic field is strong
with the cyclotron frequency
that is larger than the photon frequency $\nu_B \gg \nu$
and the wave electric vector is
nearly perpendicular to the magnetic field
with $\sin\theta_{E} \approx \theta_{E} \ll 1$
\citep[the inner product of unit vectors
along the magnetic field and the wave electric field;][]{Canuto+71,Kumar_Lu20}.
The propagation of FRB photons could be possible if
the FRB photons are generated with extreme beaming
$\theta_E < \tau_C^{-1/2} \sim 10^{-11} r_7^{5/2}$.
We do not consider this case in this Letter.

The plasma frequency
$\nu_p \sim \left({e^2 n_{\pm}}/{\pi m_e}\right)^{1/2}
  \sim 3\times 10^3\,{\rm GHz}\,
  \Gamma_{d,0.4}^{13/6} r_{7}^{-3/2}$
is also higher than the photon frequency \citep{Yamasaki+19}.
The optical depth to free-free absorption may be also high.
These constraints are also mitigated if
particle motion is restricted by the strong magnetic field \citep{Kumar+17}.
In any case, the system is optically thick
for FRB photons.

\subsection{Breakout of FRB Photons from the $e^{\pm}$ Outflow}\label{sec:breakout}

FRB photons from the magnetosphere are observable if they push aside
and break out of the surrounding $e^{\pm}$ outflow
via induced Compton scatterings.
The FRB energy is wasted into pushing the $e^{\pm}$ outflow.
In this Letter, we adopt a simple criteria for the breakout:
the work done by the FRB photons on the $e^{\pm}$
is less than the FRB energy.

The work done on the $e^{\pm}$ is estimated as follows.
Let us consider the comoving frame of the $e^{\pm}$ outflow.
The propagation speed of the head of the FRB photons
should be close to light speed $c$
in order for the breakout within the dynamical time.
The pushed $e^{\pm}$ is heated up and
the wasted energy per volume is at least $\sim n'_{\pm} m_e c^2$.
However, the actual wasted energy is much more
because of the Compton drag (or cooling) by the X-ray bursts on the $e^{\pm}$
\citep[see also][]{Cordes+16,Katz+20}.
The Compton cooling carries away energy $\sim c \sigma_T u'_X t'_{dyn}$
from a heated electron (or positron)
as the $e^{\pm}$ heating generally
continues for the dynamical time $t'_{dyn}=r/c\Gamma_{\pm}$.\footnote{
  There is a configuration in which
  the heating time is much less than $t'_{dyn}$.
  However, this is not general
  because there is a relative velocity
  between the FRB emission region and the $e^{\pm}$ outflow.
}
This is larger than the rest mass energy $m_e c^2$
as shown in Eq.~(\ref{eq:drag}).
Therefore the wasted energy per volume is
$\sim n'_{\pm} c t'_{dyn} \sigma_T u'_{X}$.

This wasted energy density
should be less than the energy density of the FRB photons,
$u'_{\rm FRB}=L_{\rm FRB}/4\pi r^2 c \Gamma_{\pm}^2$, as
\begin{eqnarray}
  u'_{\rm FRB} > n'_{\pm} c  t'_{dyn} \sigma_T u'_X= \tau_T u'_X
\end{eqnarray}
where $\tau_T$ is the Thomson optical depth in Eq.~(\ref{eq:Thomson}).
This results in a simple breakout criteria with Eq.~(\ref{eq:npm(r)}),
\begin{eqnarray}
  1 &<& \frac{L_{\rm FRB}}{\tau_T L_X}
  =\frac{L_{\rm FRB}}{L_X}\frac{\Gamma_{\pm}^2}{\Gamma_d^{13/3}}
  \left(\frac{r}{R}\right)^2
  \nonumber\\
  &\sim& 2 \times 10^{-2}\, L_{{\rm FRB},38.6} L_{X,41}^{-1}
  \Gamma_{d,0.4}^{-10/3} \Gamma_{\pm}^{2} r_7^{2},
  \label{eq:breakout}
\end{eqnarray}
where $L_{\rm FRB}=4 \times 10^{38}$ erg s$^{-1}\,L_{{\rm FRB},38.6}$
is the isotropic FRB luminosity
\citep{Bochenek+20,CHIME/FRB+200428}.
Therefore, the breakout is possible if
the emission radius $r_{\rm FRB}$ is larger than
\begin{eqnarray}
  r_{\rm FRB} > 7 \times 10^{7}\,{\rm cm}\,
  L_{{\rm FRB},38.6}^{-1/2} L_{X,41}^{1/2}
  \Gamma_{d,0.4}^{5/3} \Gamma_{\pm}^{-1},
  \label{eq:rFRB>}
\end{eqnarray}
where the $e^{\pm}$ Lorentz factor $\Gamma_{\pm}$ is determined
by the Compton drag in Eqs.~(\ref{eq:beta}) or (\ref{eq:G*})
and basically mildly relativistic.
The upper limit on the emission radius is determined by the energetics
$u_{\rm FRB} > B^2/8\pi$ as
\begin{eqnarray}
  r_{\rm FRB} < 1\times 10^{9}\,{\rm cm}\,B_{p,14.3}^{1/2} L_{{\rm FRB},38.6}^{-1/4}.
  \label{eq:rFRB<}
\end{eqnarray}
Note that the breakout condition $L_{\rm FRB}>\tau_T L_X$
  in Eq.~(\ref{eq:breakout}) is applicable even if
  the pair density is determined by a different mechanism
  from Sec~\ref{sec:TEFB}.

\begin{figure}
  \begin{center}
    \includegraphics[width=\linewidth]{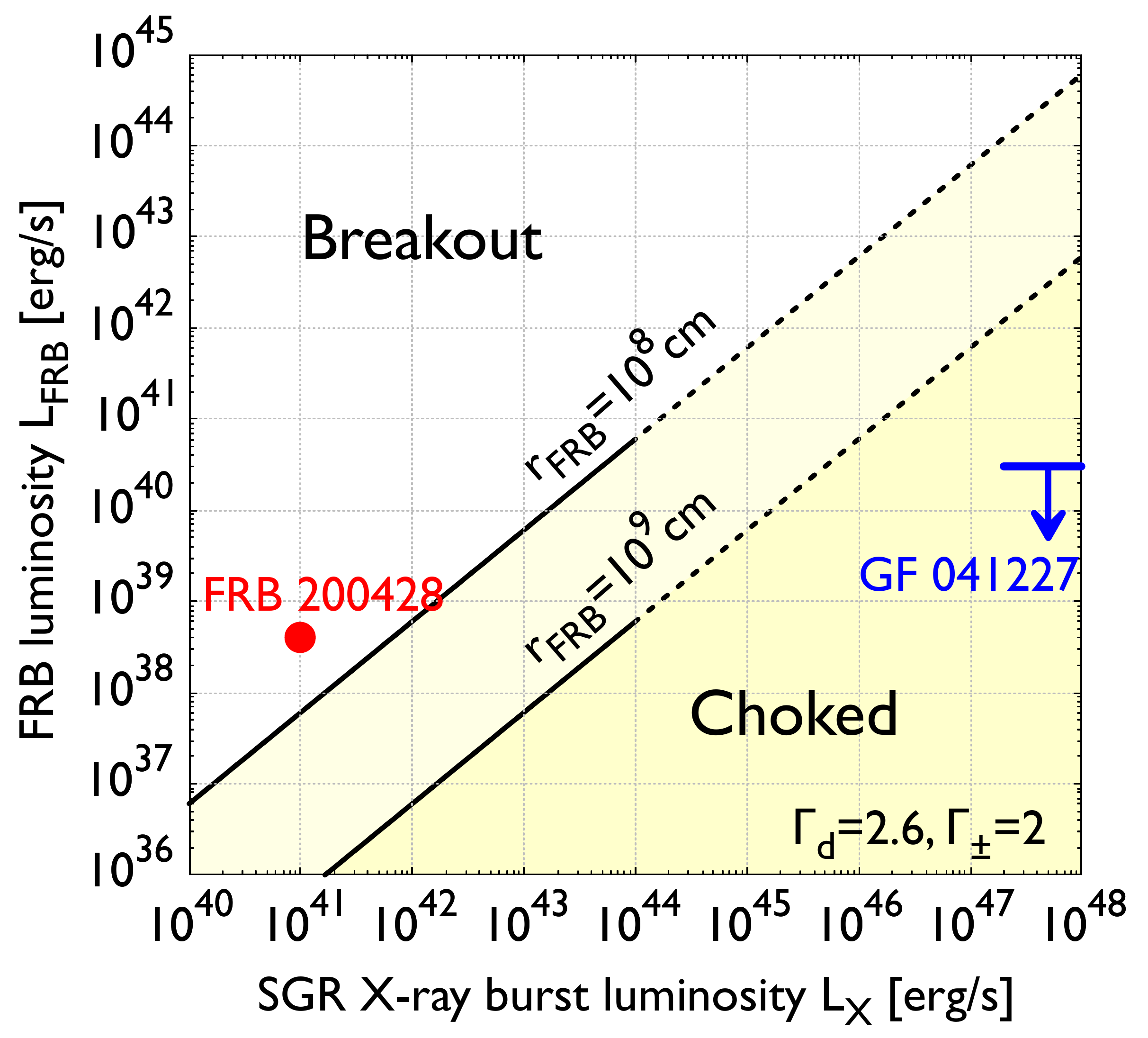}
  \end{center}
  \caption{
    Extrapolation of the breakout condition in Eq.~(\ref{eq:breakout})
    on the plane of the FRB and X-ray burst luminosities
    for the cases of emission radii
    $r_{\rm FRB}=10^{8}$ cm and $r_{\rm FRB}=10^{9}$ cm
    with $\Gamma_d=2.6$ and $\Gamma_{\pm}=2$.
    FRB 200428 can break out of the $e^{\pm}$ outflow associated with
    the X-ray bursts if the emission radius is larger than
    a few tens of neutron star radii in Eq.~(\ref{eq:rFRB>}).
    For the giant flare of 2004 December 27 from SGR 1806-20
    \citep{Hurley+05,Terasawa+05},
    a FRB weaker than the radio limit \citep{Tendulkar+16}, if any,
    would be choked by the $e^{\pm}$ outflow.
    The extrapolation may not be reliable
    for $L_{X} \gtrsim 10^{44}$ erg s$^{-1}$
    because the trapped fireball size $\ell_{X}$ in Eq.~(\ref{eq:lx})
    becomes comparable to the NS radius.
  }
  \label{fig:LLx}
\end{figure}

Figure~\ref{fig:LLx} extrapolates the breakout condition
in Eq.~(\ref{eq:breakout})
to the other FRB and X-ray burst luminosities
in the cases of emission radii
$r_{\rm FRB}=10^{8}$ cm and $r_{\rm FRB}=10^{9}$ cm
with $\Gamma_d=2.6$ and $\Gamma_{\pm}=2$.
We can see that the breakout condition requires
brighter FRBs for brighter X-ray bursts.
Further implications will be discussed in Sec.~\ref{sec:summary}.

\section{Summary and Discussions}\label{sec:summary}

We show that the $e^{\pm}$ outflow is accompanied by
the SGR X-ray bursts with high cutoff energy $T_{\rm cut} \sim 80$ keV
by modeling the trapped--expanding fireball.
The FRB photons can not propagate in the $e^{\pm}$ outflow
due to induced Compton scatterings, but can break it out
if the emission radius is larger than a few tens of NS radii
in Eq.~(\ref{eq:rFRB>}) and Fig.~\ref{fig:LLx}.
The breakout condition also puts upper limits to
X-ray counterparts of cosmological FRBs
\citep[see][]{Scholz+17,Scholz+20}.

The FRB pulse widths
\citep[$\sim 0.6$ ms;][]{Bochenek+20,CHIME/FRB+200428}
are shorter than the delay
\citep[$\sim 6.5\pm 1.0$ ms;][]{Mereghetti+20}
and width \citep[$\sim 3$ ms;][]{Li+20}
of the X-ray bursts.
This suggests that the X-rays are trapped by the trapped fireball,
yielding the comparable times for the delay and width,\footnote{
  Note that this delay is about a single energy injection.
  There are likely multiple or extended energy injections,
  and the first one, which produces the onset of the X-ray bursts,
  is not associated with an FRB as observed.
}
while the intrinsic timescale of the energy generation
is shorter than the trapping time,
and the energy generation radius is less than
$0.6\,{\rm msec} \times c \sim 2\times 10^{7}$ cm.
This is below
the FRB emission radius limited 
by the breakout condition in Eq.~(\ref{eq:rFRB>}),
requiring energy transfer, e.g., by MHD waves.

Other X-ray bursts are not associated with FRBs
down to eight orders of magnitude fainter than FRB 200428
\citep{Lin+20}.
One possibility is that 
the $e^{\pm}$ outflow from an expanding fireball
could be essential for the coherent radio emission:
in the other X-ray bursts with low $T_{\rm cut}$,
the surface of the trapped fireball is transparent in Fig.~\ref{fig:tau}
and the expanding fireball is not launched.
Although the $e^{\pm}$ outflow
is less energetic than the FRB in Eq.~(\ref{eq:Lpm}),
it could affect the coherent condition of the FRB emission.\footnote{
  In the binary comb model \citep{Ioka+20},
  aurora particles could change the coherent condition
  \citep[see also][for another possible particle]{Dai20}.
  There might be several channels to FRBs.
}
Another possibility is that
an open field line could be necessary for transferring the MHD waves, or
faint FRBs are choked by
the $e^{\pm}$ outflow associated with the X-ray bursts
in Fig.~\ref{fig:LLx}.

No FRB was detected at the giant flare 2004 December 27
from SGR 1806-20
\citep{Hurley+05,Terasawa+05}
with a radio limit of $110$ MJy ms at 1.4 GHz \citep{Tendulkar+16}.
A FRB similar to FRB 200428, if any, is 
choked by the $e^{\pm}$ outflow as in Fig.~\ref{fig:LLx},
while a very bright FRB can break it out.

\citet{Kirsten+20} detected
two radio bursts with $112\pm 22$ Jy msec and $24\pm 5$ Jy ms,
$4$--$5$ orders of magnitude fainter than FRB 200428.
Accompanying X-ray bursts are expected to be faint from
the breakout condition in Eq.~(\ref{eq:breakout}) and Fig.~\ref{fig:LLx},
consistent with the non-detection.
Very recently CHIME/FRB detected three radio bursts
  with $900\pm 160$, $9.2\pm 1.6$, and $6.4 \pm 1.1$ Jy ms
  \citep{ATel14074,ATel14080}
  without gamma-ray counterparts \citep{ATel14087}.
  This is also consistent with
  Eq.~(\ref{eq:breakout}) and Fig.~\ref{fig:LLx}.

Further studies are needed to better understand the entire breakout process,
such as shock structure,
motion of heated $e^{\pm}$ along magnetic fields,
emission from the heated $e^{\pm}$, and so on,
as well as baryon loading to the fireball.

If the energy release is caused by magnetic reconnection,
similar energies are ejected in the opposite directions,
so that the outflow is as energetic as the X-ray bursts
\citep{Yamasaki+20b,Yamasaki+19,Yuan+20}.
The energy is many orders of magnitude larger than
that calculated in this Letter in Eq.~(\ref{eq:Lpm}).
Hence, completely different afterglows or nebulae are expected.
Note that for the radio afterglow of
the giant flare on 2004 December 27 from SGR 1806-20,
the minimum energy is smaller than the flare energy
\citep{Cameron+05,Gaensler+05,Nakar+05}.
In contrast, the ratio is unity for gamma-ray bursts.
This implies that the outflow is less energetic than the flare or X-ray bursts,
but the definite conclusion requires further studies.

\acknowledgments

The author would like to thank
the referee for valuable comments, and
H.~Hamidani,
W.~Ishizaki,
K.~Kashiyama,
S.~Kisaka,
K.~Murase,
K.~Takahashi,
T.~Wada,
S.~Yamasaki,
and B.~Zhang
for useful discussions.
This work is partly supported by
JSPS KAKENHI Nos. 20H01901, 20H01904, 20H00158, 18H01213, 18H01215, 17H06357, 17H06362, and 17H06131.
Discussions during the YITP workshop YITP-T-19-04 and YKIS2019
are also useful for completing this work.

\bibliography{ref}{}
\bibliographystyle{aasjournal}



\end{document}